\documentstyle[epsfig,12pt]{article}

\renewcommand{\thefootnote}{\fnsymbol{footnote}}

\def\Journal#1#2#3#4{{#1} {\bf #2}, #3 (#4)}

\def\PLB{{\em Phys. Lett.}  B}
\def\PRL{\em Phys. Rev. Lett.}
\def\PRD{{\em Phys. Rev.} D}

\begin{document}

\epsfysize3cm
\epsfbox{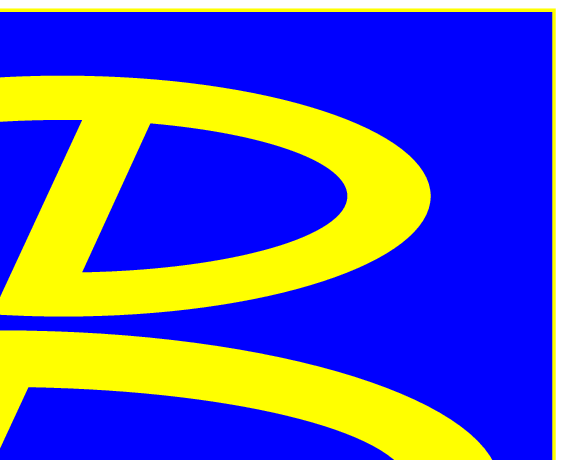}    

\vskip -3cm
\noindent
\hspace*{12cm}KEK Preprint 2001-20  \\
\hspace*{12cm}Belle Preprint 2001-6 \\

\vskip 2cm

\begin{center}

{\Large\bf 
Study of Rare $B$ Meson Decays at Belle
}

\vskip 0.5cm

Toru Iijima

\vskip 0.2cm
{\it 
High Energy Accelerator Research Organization (KEK), 1-1 Oho, Tsukuba,
Japan \\
E-mail: toru.iijima@kek.jp}   

\vskip 0.2cm
for the Belle Collaboration

\end{center}

\vskip 1.0cm

\begin{abstract}
In this paper, we briefly review results on rare decays of $B$ mesons
from the Belle experiment, based on 10.4\,fb$^{-1}$ data collected
on the $\Upsilon$(4S) resonance.
The topics include measurements of Cabibbo suppressed decays $B \to
D^{(*)} K^{(*)}$, charmless two-body decays $B \to \pi\pi$, $K\pi$,
$KK$ and $B \to \eta^{'} h\,(h = K^+, \pi^+, K^0)$, and a search for
flavor changing neutral current decays $B \to K^{(*)} \ell^{+}
\ell^{-}$.
\end{abstract}

\vskip 1.0cm

\begin{center}
{\sl
To appear in the Proceedings of the \\
4th International Workshop on B Physics and CP Violation (BCP4) \\
February 19--23, 2001, Ise-Shima, Japan.
}
\end{center}

\clearpage

\renewcommand{\thefootnote}{\alph{footnote}}
\setcounter{footnote}{0}

\section{Introduction}
\label{sec:intro}
Rare decays of the $B$ meson proceed via a variety of diagrams.
They provide a unique opportunity to study the Kobayashi-Maskawa
matrix elements, $|V_{cb}|$, $|V_{ub}|$, $|V_{ts}|$ and $|V_{td}|$,
as well as their relative phases.
In addition, decays involving loop diagrams are interesting because
of their sensitivity to effects of new physics.
Such rare $B$ meson decays can be studied in detail at the recently
commissioned $e^+e^-$\,$B$-factories, KEKB and PEP-II.

By the end of December 2000, the Belle experiment~\cite{belle} at
the KEKB accelerator~\cite{kekb} had accumulated 
10.4\,fb$^{-1}$ of data on the $\Upsilon$(4S) resonance, corresponding
to $11.1 \times 10^6$\, $B \bar{B}$ pairs.
Based on this data set, results for the following rare decays are
briefly reviewed in this paper;
1) Cabibbo suppressed decays $B \to D^{(*)} K^{(*)}$, 
2) charmless two-body decays $B \to \pi\pi$, $K\pi$, $KK$,
3) charmless decays involving the $\eta^{'}$ meson
$B \to \eta^{'} h\,(h = K^+, \pi^+, K^0)$, and
4) a search for flavor changing neutral current (FCNC) decays 
$B \to K^{(*)} \ell^{+} \ell^{-}$.
Results for additional rare decay modes can be found in other
contributions to this conference~\cite{others}.

In all the analyses, $B$ candidates are reconstructed using
the beam constrained mass, $m_{bc}=\sqrt{E_{\rm beam}^2-p_B^2}$, and
the energy difference, $\Delta E=E_B-E_{\rm beam}$.
Here, $E_{\rm beam} \equiv \sqrt{s}/2\simeq 5.290$ GeV, and $p_B$ and
$E_B$ are the momentum and energy of the reconstructed $B$ in the
$\Upsilon$(4S) rest frame, respectively
\footnote{In the analysis for the Cabibbo suppressed decays,
described in Sec.~\ref{sec:dk}, the beam constrained mass is
calculated in the laboratory frame as 
$m_{bc} = \sqrt{(E^{lab}_{B})^{2} - |\vec{p}_B^{\,lab}|^{2}}$.
Here $\vec{p}_B^{\,lab}$ is the $B$ candidate's laboratory momentum 
vector and 
$E^{lab}_{B} = {1\over E_{ee}}(s/2 + \vec{P}_{ee} \cdot 
\vec{p}_B^{\,lab})$, where $s$ is square of the center of mass 
energy, and $\vec{P}_{ee}$ and $E_{ee}$ are the laboratory  
momentum and energy of the $e^+e^-$ system, respectively.}.
Normally we compute $\Delta E$ assuming a $\pi$ mass for each charged
particle from $B$ decays.
This shifts $\Delta E$ by about $-50$\,MeV for each $K^{\pm}$
meson.
Other common analysis techniques, the particle identification (PID)
cut and the $q\bar{q}$ background suppression cut, are described in
appendices.
Throughout this paper, the inclusion of charge conjugate states is
implied, except for the direct $CP$ violation measurement described in
Sec.~\ref{sec:hh}.
The first and second errors in results represent statistical and
systematic errors, respectively.

\section{Cabibbo Suppressed Decays}
\label{sec:dk}
Direct $CP$-violating asymmetries due to interference between 
$b \to c$ and $b \to u$ transition amplitudes in the Cabibbo
suppressed $B^- \to D^0 K^-$ decay is a theoretically clean way to
determine the $\phi_3$ angle of the unitarity
triangle~\cite{dk:gronau}.
As a first step of this program, we have searched for the Cabibbo
suppressed processes  $B^- \to D^{*0} K^-$, $\bar{B}^0 \to D^{*+}
K^-$, $\bar{B}^0 \to D^{+} K^-$ as well as $B^- \to D^0 K^-$.
In the tree level approximation, their branching fractions are
related to those of the Cabibbo favored $B \to D^{(*)} \pi^-$
counterparts by 
\begin{equation}
R \equiv
\frac
{{\cal B}(B \to D^{(*)} K^-)}{{\cal B}(B \to D^{(*)} \pi^-)}
\simeq \tan^{2} \theta_{C} (f_{K}/f_{\pi})^2 \simeq 0.074,
\label{eq:dkratio}
\end{equation}
using the Cabibbo angle $\theta_{C}$ and the meson decay constants
$f_{K(\pi)}$
\footnote{
The relation assumes the validity of factorization and flavor-$SU(3)$
symmetry.
}.
The only Cabibbo suppressed decay observed to date is $B^- \to D^0
K^-$, reported by CLEO~\cite{dk:cleo} with $R = 0.055 \pm 0.014 \pm
0.005$.

In the analysis, $D$ mesons are reconstructed using the decay
modes, $D^0 \to K^- \pi^+$, $K^- \pi^+ \pi^0$, $K^- \pi^+ \pi^+
\pi^-$ and $D^+ \to K^- \pi^+ \pi^+$, $K_S^0 \pi^+$, $K_S^0 \pi^+
\pi^+ \pi^-$, $K^- K^+ \pi^+$.
For $D^{*0}$ and $D^{*+}$ reconstruction, $D^{*0} \to D^0 \pi^0$ and
$D^{*+} \to D^0 \pi^+$, $D^+ \pi^0$ decays are used.
If multiple entries are found in one event, the best candidate 
is selected based on a $\chi^2$ determined from the differences
between measured and nominal masses of $m_D$, $m_{bc}$ and, when
appropriate $m_{D^*} -m_D$ and $m_{\pi^0}$ ($\pi^0$ from the $D^{*}$
decays).
To suppress the much more abundant $B \to D^{(*)} \pi^-$ decays, a
relatively tight PID cut ${\cal R}_{K} > 0.8$, which gives 76.5\% 
efficiency for $K$ tracks and 2\% $\pi \to K$ fake rate, is applied
for the prompt hadron track from the $B$ decay (see Appendix A).

Figure~\ref{fig:dk1} shows the $\Delta E$ distributions for all $B \to
D^{(*)} \pi^-$ (${\cal R}_{K}<0.8$) and $B \to D^{(*)} K^-$ (${\cal
R}_{K}>0.8$) enriched samples.
In all the $B \to D^{(*)} K$ channels, signal peaks are clearly seen
at $\Delta E = -49$\,MeV.
Table~\ref{tbl:dk1} summarizes the results.
The $R$ ratios, defined in Eq.(\ref{eq:dkratio}), are calculated from
the extracted yields of 
$B \to D^{(*)} \pi^-$ and $B \to D^{(*)} K^-$ 
decays by taking into account the difference in the reconstruction
efficiencies. 
In all cases, the $R$ ratios are consistent within errors with the
expectation in Eq.(\ref{eq:dkratio}).    
These are the first observations of the $B \to D^+ K^-$, $D^{*0} K^-$ 
and $D^{*+} K^-$ decay processes
\footnote{
Results on $B \to D^{(*)} K$ modes were updated after the conference.
}.

\begin{figure}[p]
\begin{center}
  \epsfysize 10.0cm
  \epsfbox{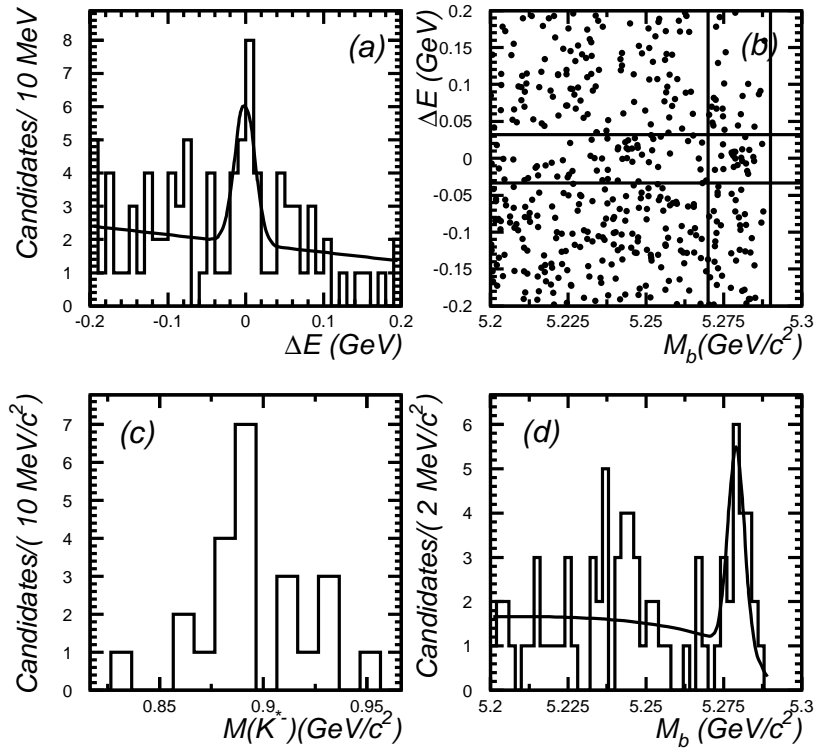}
  \end{center}
\caption{
Plots which show evidence for the $B^- \to D^0 K^{*-}$ decay;
Distributions of {\bf (a)} $\Delta E$ for the $m_{bc}$ signal region,
{\bf (b)} $m_{bc}$-$\Delta E$, {\bf (c)} $K_S^0 \pi^-$ invariant mass
for events in the $m_{bc}$-$\Delta E$ signal box and {\bf (d)}
$m_{bc}$ for the $\Delta E$ signal region. The solid lines in {\bf
(b)} indicate the $m_{bc}$ and $\Delta E$ signal regions.
}
\label{fig:dk2}
\end{figure}

We have also searched for the $B^- \to D^0 K^{*-}$ decay mode.
The $K^{*}$ mesons are reconstructed using the clean $K_S^0 \pi^-$
channel.
The first observed evidence of this decay mode is shown in
Fig.~\ref{fig:dk2}.
The signal yield is $15.0 \pm 4.6$ with a statistical
significance of $4.6\,\sigma$.
The branching fraction normalized to that of $B^- \to D^0 \pi^-$
is found to be $0.116 \pm 0.036 \pm 0.015$ (preliminary)
\footnote{
The ratio similar to Eq.(\ref{eq:dkratio}), $R \equiv {\cal B}(B^- \to
D^0 K^{*-})/{\cal B}(B^- \to D^0 \rho^-)$, is not presented here
because our analysis for the Cabibbo favored counterpart
$B^- \to D^0 \rho^-$ has not been completed yet.
}.

\section{$B \rightarrow \pi\pi, K\pi, KK$ Decays}
\label{sec:hh}
Branching fraction measurements of the $B \to \pi\pi$, $K \pi$ and
$KK$ decays are an important first step toward indirect and direct
$CP$ violation studies with the $\pi\pi$ and $K\pi$ modes,
which are related to the angles $\phi_2$ and $\phi_3$ of the unitarity
triangle, respectively.
Moreover, assuming isospin and $SU(3)$ symmetry, recent theoretical
work suggests that branching fraction of these modes can be used to 
extract or place limits on the angles $\phi_2$ and
$\phi_3$~\cite{hh:phys}.

We have analyzed the data for the $B^0 \to \pi^+\pi^-$, $K^+\pi^-$,
$K^+K^-$, $K^0\pi^0$ and $B^+ \to \pi^+\pi^0$, $K^+\pi^0$, $K^0\pi^+$
decays.
The Belle detector is equipped with a high momentum PID system, which
gives clear separation between the charged $\pi$ and $K$ mesons in
these final states.
The $\pi$ and $K$ mesons are distinguished by the PID cut
${\cal R}_{\pi(K)} > 0.6$, which gives 92.4\% efficiency and 4.3\%
fake rate (true $\pi$ fakes $K$) for $\pi$ mesons and 84.9\%
efficiency and 10.4\% fake rate (true $K$ fakes $\pi$) for
$K$ mesons, respectively (see Appendix A).
The dominant background from the continuum $q\bar{q}$ process is
suppressed using the likelihood ratio method described in Appendix B.

\begin{figure}[p]
  \begin{center}
  \epsfysize 10.0cm
  \epsfbox{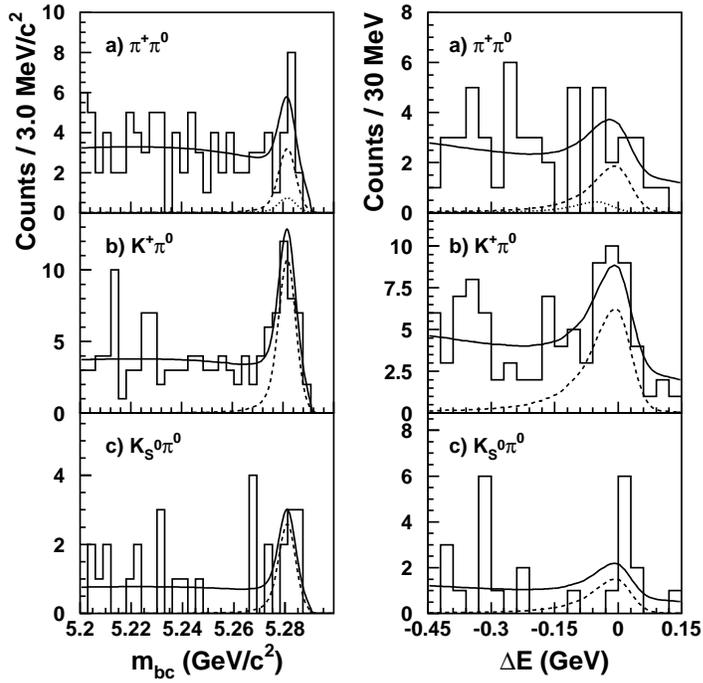}
  \end{center}
\caption{{}
	The $m_{bc}$ (left) and $\Delta E$ (right) distributions for
	$B^+\rightarrow$ a) $\pi^+\pi^0$, b) $K^+\pi^0$ and c) $K^0_S\pi^0$.
	For the $K^+\pi^0$, $K$ mass is assumed for the charged particle.
	The projection of the two dimensional fit onto each variable and
	its signal component are shown by the solid and dashed curve,
	respectively. 
	In the $\pi^+\pi^0$ fit, the cross talk from $K^+\pi^0$
	is indicated by the dotted curve.
}  
\label{fig:hh2}
\end{figure}

Figure \ref{fig:hh1} shows $m_{bc}$ and $\Delta E$ distributions for
the $\pi^+\pi^-$, $K^+\pi^-$ and $K^0_S\pi^+$ modes.
Each distribution is fitted to a Gaussian signal on top of a
background function, which is modeled by the ARGUS background
function~\cite{hh:argus} for $m_{bc}$ and by a linear function for
$\Delta E$.
The $\Delta E$ fit results are used to determine the signal yields.
For the $\pi^+\pi^-$ and $K^+\pi^-$ modes, the cross-talk from each mode
is not negligible, and thus the fits treat the size of both components
as free parameters.
The cross talk found in these fits are consistent with expectations.
Figure \ref{fig:hh2} shows the $m_{bc}$ and $\Delta E$ projections
for the $\pi^+\pi^0$, $K^+\pi^0$ and $K^0_S\pi^0$ modes.
For these modes, since the $\Delta E$ distribution has a long tail,
a two-dimensional fit is applied to the $m_{bc}$ and $\Delta E$
distributions.
For the $\pi^+\pi^0$ mode, since the cross talk from $K^+\pi^0$ is
significant and separated by less than 1$\sigma$ in $\Delta E$, the
fit includes the expected $K^+\pi^0$ component.

Results of the branching fraction measurements are summarized in Table
\ref{tbl:hh1}
\footnote{
Results shown in Table~\ref{tbl:hh1} were updated after the conference.
}.
The charge averaged branching fractions for $B \to \pi^+\pi^-$,
$K^+\pi^-$, $K^+\pi^0$, $K^0\pi^+$, and $K^0\pi^0$ are measured with
statistically significant signals.
Our results are consistent with other
measurements~\cite{hh:cleo}$^{,}$\cite{hh:babar}, confirm that
${\cal B}(B^0 \to K^+\pi^-)$ is larger than ${\cal B}(B^0 \to
\pi^+\pi^-)$, and indicate that ${\cal B}(B^+ \to h^+\pi^0)$ and
${\cal B}(B^0 \to K^0\pi^0)$ is larger than expected in relation to
the $B^0 \to h^+\pi^-$  and $B^+ \to K^0\pi^+$ modes based on isospin
or penguin dominance~\cite{hh:phys}.

We have also searched for direct $CP$ violation in $B \to K^{\pm}
\pi^{\mp}$ and $B \to K^{\pm} \pi^{0}$ modes.
With the high momentum PID at Belle, the dilution of the $CP$
asymmetry due to double misidentification is
as small as $\sim 0.5\%$. 
Inherent asymmetries in the Belle detector are determined to be less
than $2\%$ based on the yield difference between $\bar{D}^0 \to K^+
\pi^-$ and $D^0 \to K^- \pi^+$ decays.
Figure~\ref{fig:hh3}-a) and b) shows the $\Delta E$ distributions for
the $K^- \pi^+$ and $K^+ \pi^-$ final states.   
The signal yields, extracted with the same procedure as the branching
fraction measurement, are
$27.7^{+6.8}_{-6.1}$ for $B^0 \to K^- \pi^+$ and  
$25.4^{+7.0}_{-6.3}$ for $\bar{B}^0 \to K^+ \pi^-$, giving a partial
rate asymmetry
\footnote{
Here the partial rate asymmetry ${\cal A}_{cp}$ is defined as,
${\cal A}_{cp} \equiv \frac{N(\bar{B} \to \bar{f}) - N(B \to f)}
                           {N(\bar{B} \to \bar{f}) + N(B \to f)}$.  
} ${\cal A}_{cp}(K^{\pm}\pi^{\mp}) = 0.043 \pm 0.175 \pm 0.021$ and 
its 90\% confidence level interval 
$-0.26 < {\cal A}_{cp}(K^{\pm}\pi^{\mp}) < 0.35$.
Figure~\ref{fig:hh3}-c) and d) show the same distributions for the
$K^- \pi^0$ and $K^+ \pi^0$ final states.
The yields are 
$18.3^{+5.6}_{-4.9}$ for $B^- \to K^- \pi^0$ and  
$17.6^{+5.5}_{-4.8}$ for $B^+ \to K^+ \pi^0$, giving 
${\cal A}_{cp}(K^{\pm} \pi^0) = 0.019^{+0.219}_{-0.191}$
(only statistical error is shown for this mode).
It should be noted that these results on $A_{cp}$ are preliminary.

\section{$B \rightarrow \eta^{'}h$ Decays}
\label{sec:etaph}
Measurements by CLEO~\cite{etapk:cleo} and BABAR~\cite{hh:babar} 
indicate that the branching fraction for $B \to \eta^{'} K$ is
significantly larger than theoretical expectations.
Here, we present preliminary results for the $B^+ \to \eta^{'} K^+$,
$\eta^{'} \pi^+$ and $B^0 \to \eta^{'} K^0$ decay modes.
The $\eta^{'}$ mesons are reconstructed using $\eta^{'} \to \eta \pi
\pi$, $\eta \to \gamma \gamma$ and $\eta^{'} \to \rho \gamma$ decays.
The $q \bar{q}$ background is suppressed by a likelihood ratio
cut similar to that used for the $B \to \pi\pi$, $K\pi$, $KK$
analysis.
The likelihood contains $SFW$ and $\cos \theta_B$ (see Appendix B).
The prompt charged $K/\pi$ from the $B$ decay is separated by the
PID cut ${\cal R}_{\pi(K)} > 0.6$.

\begin{figure}[p]
\begin{center}
\epsfysize 10.0cm
\epsfbox{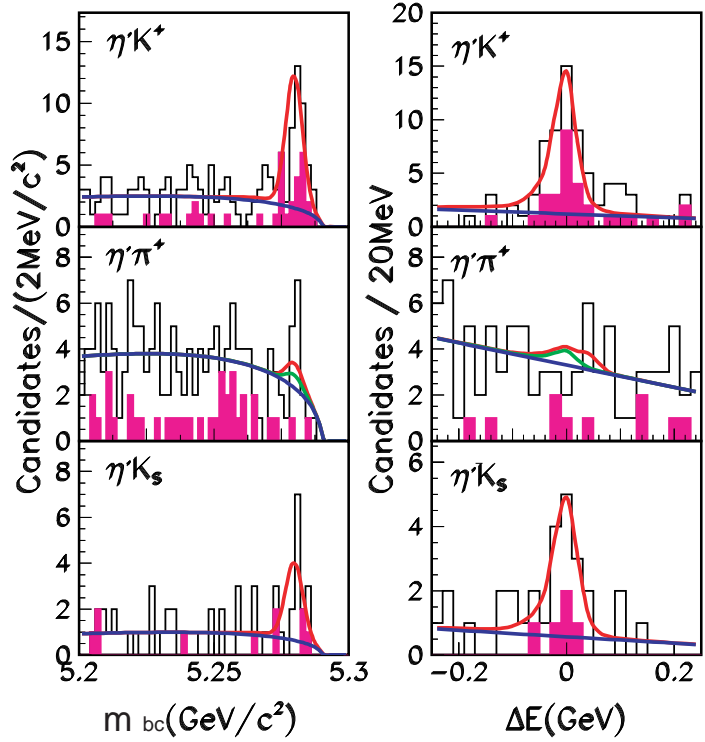}
\end{center}
\caption{{}
The $m_{bc}$ and $\Delta E$ distributions for
(top) $B^+ \to \eta^{'} K^+$,
(middle) $B^+ \to \eta^{'} \pi^+$, and
(bottom) $B^0 \to \eta^{'} K_S^0$
mode.
The hatched histograms show the contribution from 
$\eta^{'} \to \rho \gamma$ sub-decay mode.
The solid lines show the fits with a Gaussian signal on top of
a background function, which is the ARGUS background function for
$m_{bc}$ and a linear function for $\Delta E$.
The fits for $B^+ \to \eta^{'} \pi^+$ include the expected   
$B^+ \to \eta^{'} K^+$ component.}
\label{fig:etapk}
\end{figure}

Figure~\ref{fig:etapk} shows the $m_{bc}$ and $\Delta E$ distributions
for the signal candidates.
By combining results for the two sub-decay modes of the $\eta^{'}$
meson, the observed signals have statistical significances of $10.9
\sigma$ and $5.9 \sigma$ for the $\eta^{'} K^+$ and $\eta^{'} K^0$ 
modes, respectively.
Their branching fractions are found to be ${\cal B}(B^+ \to 
\eta^{'} K^+) = (6.8^{\ +1.3\ +0.7}_{\ -1.2\ -0.9}) \times 10^{-5}$
and ${\cal B}(B^0 \to \eta^{'} K^0) = (6.4^{\ +2.5\ +1.0}_{\ -2.0\
-1.1}) \times 10^{-5}$.
In the $\eta^{'} \pi^+$ mode, no statistically significant signal
is seen, and we set an upper limit of 
${\cal B}(B^+ \to \eta^{'} \pi^+) < 1.2 \times 10^{-5}$ (90\% C.L.).
These results are consistent with the previous measurements.

\section{Search for $B \to K^{(*)} \ell^{+} \ell^{-}$ Decays}
\label{sec:kll}
The FCNC decay, $B \to K^{(*)} \ell^{+} \ell^{-}$, proceeds via loop
diagrams, and therefore is sensitive to new physics such as charged
Higgs and SUSY.
We have searched for the FCNC decays,
$B^0 \to K^{*0} \ell^{+} \ell^{-}$,  
$B^+ \to K^{*+} \ell^{+} \ell^{-}$,  
$B^0 \to K^{0} \ell^{+} \ell^{-}$,  
$B^+ \to K^{+} \ell^{+} \ell^{-}$,  
in both electron and muon channels.

The $K^0$ mesons are reconstructed using $K_S^0 \to \pi^+ \pi^-$ decays. 
The $K^{*}$ mesons are reconstructed using $K^{*0} \to K^+ \pi^-$,
$K_S^0 \pi^0$ and $K^{*+} \to K_S^0 \pi^+$, $K^+ \pi^0$ decays.
Electrons with $p_{lab} > 0.5$\,GeV/$c$ and muons with $p_{lab} >
1.0$\,GeV/$c$ are selected with pion fake rates expected at a level of 
0.3\% and 1.7\%, respectively.

Background from $q \bar{q}$ processes is suppressed by a likelihood
ratio cut (see Appendix B), where the likelihood contains $\cos \theta_B$, 
$\cos \theta_{da}$ and a Fisher discriminant constructed from virtual
calorimeter variables\cite{kll:vcal} and the normalized second
order Fox-Wolfram momentum $R_2$.
The largest source of $B \bar{B}$ background are events where both $B$
mesons decay into $\ell \nu X$.
This source is suppressed by cutting on a likelihood ratio based on 
visible energy and $\cos \theta_B$. 
The large background from $J/\psi \to \ell^{+} \ell^{-}$ and $\psi^{'} 
\to \ell^{+} \ell^{-}$ decays is vetoed by applying the di-lepton
mass cuts, $-0.15 < M_{ee} - M_{J/\psi(\psi^{'})} < 0.07$\,GeV/$c^2$
and $-0.10 < M_{\mu\mu} - M_{J/\psi(\psi^{'})} < 0.05$\,GeV/$c^2$. 
The di-electron mass is required to be larger than 0.1\,GeV/$c^2$ to
avoid background from $\pi^0$ Dalitz decays and $\gamma$ conversions.

\begin{figure}[p]
\begin{center}
\epsfysize 10.0cm
\epsfbox{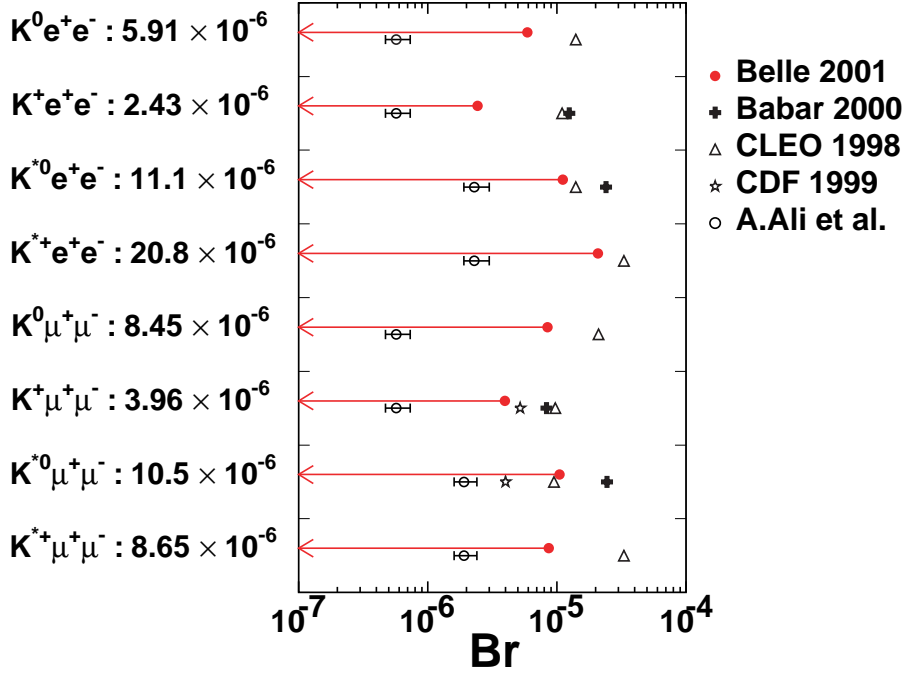}
\end{center}
\caption{{}
Upper limits on the branching fractions for each 
$B \to K^{(*)} \ell \ell$ mode (preliminary).
Our results are compared to theoretical predictions
and upper limits from previous experiments.
}
\label{fig:kll}
\end{figure}

No statistically significant signals have been observed yet, and
upper limits at 90\% confidence level are set for each mode, which are
shown in Fig.~\ref{fig:kll} along with a comparison to theoretical
expectations~\cite{kll:ali} and the results from previous
experiments~\cite{kll:cleo}$^{,}$\cite{kll:cdf}$^{,}$\cite{kll:babar}.
Our upper limits are more restrictive than those of previous
experiments, except for the $K^{*0} \mu \mu$ channel.

\section{Summary}
In this paper, we briefly review results for some $B$ meson rare
decays from the Belle experiment, based on 10.4 fb$^{-1}$ data
collected on the $\Upsilon$(4S) resonance.
They are summarized as follows,
\begin{itemize}
\item 
Cabibbo suppressed decays $B \to D^{(*)} K$ have been observed
with nearly the expected size relative to $B \to D^{(*)} \pi$ decays.
We claim the first observation of the four modes, 
$B \to D^+ K^-$, $D^{*0} K^-$, $D^{*+} K^-$ and $D^0 K^{*-}$.
\item
Our results for the $B \to \pi\pi$, $K \pi$, $KK$ modes are consistent 
with other measurements, confirm the small $\pi\pi / K\pi$ ratio, and
indicate that the $K^0 \pi^0$ mode is larger than theoretically
expected.
\item
The large branching fractions for the $B \to \eta^{'} K^+$ and
$\eta^{'} K^0$ modes are also confirmed.
\item
Improved upper limits on the FCNC decays $B \to K^{(*)} \ell^{+}
\ell^{-}$ are obtained except for the $K^{*0} \mu^+ \mu^-$ channel.
\end{itemize}
In the near future, we anticipate results on various rare $B$ meson
decays with much higher statistics and better sensitivity than ever
before achieved.

\section*{Acknowledgments}
We wish to thank the KEKB accelerator group for the excellent operation
of the KEKB accelerator.
We acknowledge support from the Ministry of Education, Culture, Sports,
Science, and Technology of Japan and
the Japan Society for the Promotion of Science;
the Australian Research Council and
the Australian Department of Industry, Science and Resources;
the Department of Science and Technology of India;
the BK21 program of the Ministry of Education of Korea and
the CHEP SRC program of the Korea Science and Engineering Foundation;
the Polish State Committee for Scientific Research under contract
No.2P03B 17017;
the Ministry of Science and Technology of Russian Federation;
the National Science Council and the Ministry of Education of Taiwan;
the Japan-Taiwan Cooperative Program of the Interchange Association;
and the U.S. Department of Energy.

\section*{Appendix A: particle identification cut}
A clear separation of charged $\pi$ and $K$ mesons is essential to
find some rare decay signals.
We combine outputs from three sub-detectors, specific ionization loss
in the central drift chamber ($dE/dx$), time-of-flight measured by
scintillation counter arrays (TOF) and light yields measured by the
aerogel Cherenkov counter arrays (ACC), into a likelihood for each
particle assumption ${\cal L}_{K(\pi)} \equiv {\cal
L}_{K(\pi)}^{dE/dx} \times {\cal L}_{K(\pi)}^{TOF} \times {\cal
L}_{K(\pi)}^{ACC}$.
The particle identification is then performed by cutting on the
likelihood ratio ${\cal R}_{K(\pi)} \equiv
{\cal L}_{K(\pi)}/({\cal L}_{K}+{\cal L}_{\pi})$.
For prompt $\pi$ and $K$ mesons from $B$ decays, TOF does not provide
useful separation, and only $dE/dx$ and $ACC$ information are used.
The PID cut efficiency and fake rate are measured using kinematically
selected $D^{*+} \to D^0 \pi^+$, $D^0 \to K^- \pi^+$ decays.

\section*{Appendix B: continuum suppression cut}
To distinguish $B\bar{B}$ events from the $q\bar{q}$ background,
Belle has developed a new event shape variable called ``Super
Fox-Wolfram''.
The usual Fox-Wolfram moments are defined as 
$H_l=\sum_{i,j}{|\vec{p_i}||\vec{p_j}|P_l(\cos\theta_{ij})}$,
where the indices $i$ and $j$ run over all final state particles,
$\vec{p}_i$ and $\vec{p}_j$ are the momentum vectors of the particle $i$
and $j$, $P_l$ is the $l$-th Legendre polynomial, and $\theta_{ij}$ is
the angle between the two particles.
In the new method, the normalized Fox-Wolfram moments,
$R_l=H_l/H_0$, are decomposed into three terms:
$R_l=R_l^{ss}+R_l^{so}+R_l^{oo}=(H_l^{ss}+H_l^{so}+H_l^{oo})/H_0$,
where the indices $ss$, $so$, and $oo$ indicate respectively that
both, one, or neither of the particles comes from a $B$ candidate.
These are combined into a six term Fisher discriminant (Super
Fox-Wolfram), defined as
$SFW=\sum^{4}_{l=1}{(\alpha_lR_l^{so}+\beta_lR_l^{oo})}$,
where $\alpha_l$ and $\beta_l$ are Fisher coefficients and $l$=2,4 for
$\alpha_l$ and $R_l^{so}$.
The terms $R_l^{ss}$ and $R_{l=1,3}^{so}$ are excluded because they
are strongly correlated with $m_{bc}$ and $\Delta E$.

\begin{figure}[p]
\begin{center}
\epsfysize 10.0cm
\epsfbox{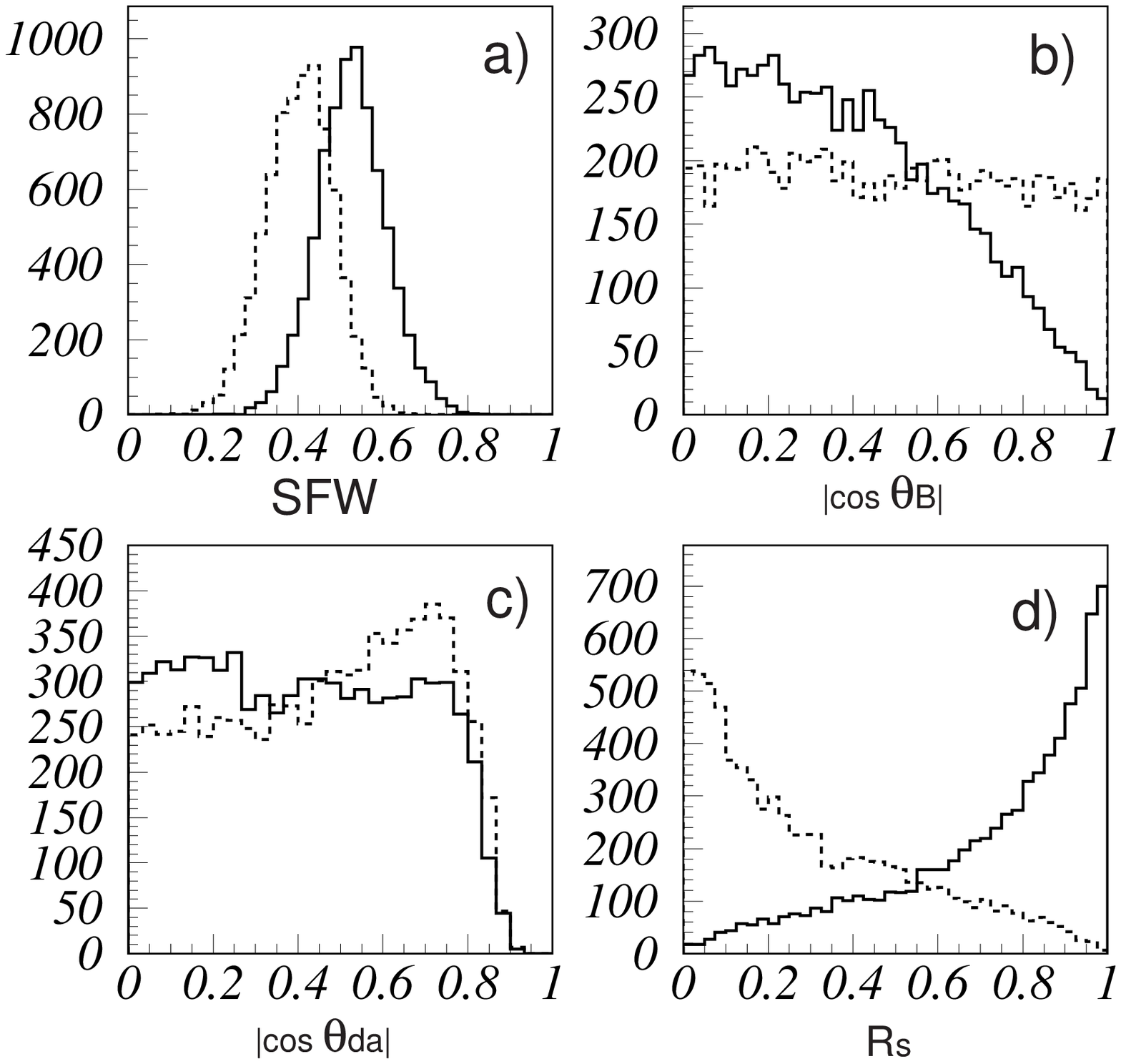}
\end{center}
\caption{{}
Distributions of the $q\bar{q}$ suppression variables in the case of
$B \to \pi^+\pi^-$, $K^+\pi^-$, $K^+K^-$ analysis; 
(A) SFW, (B) $|\cos \theta_B|$, (C) $|\cos \theta_{da}|$ and 
(D) ${\cal R}_{s}$.
Solid and dotted lines show the distributions for the $B$ decay
signal and $q\bar{q}$ background, respectively. 
}
\label{fig:appendix}
\end{figure}

We then combine different $q\overline{q}$ suppression variables into a
single likelihood,
${\cal L}_{s(q\overline{q})}=\prod_i{{\cal L}_{s(q\overline{q})}^i}$,
where the ${\cal L}_{s(q\overline{q})}^i$ denotes the
signal ($q\overline{q}$) likelihood of the suppression variable $i$,
and select candidate events by cutting on the likelihood ratio
${\cal R}_s={\cal L}_s/({\cal L}_s+{\cal L}_{q\overline{q}})$.
The variables used depend on the decay mode.
For example, Fig.~\ref{fig:appendix} illustrates the separation of the 
signal and the $q\bar{q}$ background in the case of $B \to \pi^+\pi^-$,
$K^+\pi^-$, $K^+K^-$ analysis, described in Sec.~\ref{sec:hh}, where
the likelihood contains the above $SFW$, the $B$ candidate flight
direction ($\cos \theta_B$), and the decay axis direction ($\cos
\theta_{da}$).  
By requiring ${\cal R}_s > 0.8$, 97\% of the $q\bar{q}$ background is 
removed while 48\% of the signal are kept.
In the case of $B \to \pi^+\pi^0$, $K^+ \pi^0$, $K^0\pi^0$ modes,  
the event sphericity ($S$) and the angle between the thrust axis of
the $B$ candidate and that of all the remaining particles ($\theta_T$) 
are also used.
In these modes, since the continuum background is more severe, a loose
cut on $\cos\theta_T$ is applied first.
Then $SFW$ is extended to include $\cos\theta_T$ and $S$, and this
extended $SFW$ and $\cos\theta_B$ are combined into the likelihood.


\clearpage

\begin{table}[p]
\begin{center}
\caption{
Summary of $B \to D^{(*)} K$ results. 
The extracted yields of $D^{(*)} K$ signals $N_{DK}$, 
their statistical significance $\Sigma$, and
the obtained ratios of the branching fractions $R \equiv {\cal B}(B
\to D^{(*)} K^-)/{\cal B}(B \to D^{(*)} \pi^-)$,
are shown. 
}
\label{tbl:dk1}  
\vspace{0.2cm}
\begin{tabular}{lccc}  
\hline\hline
channel  & $N_{DK}$ & $\Sigma$ & $R$ \\
\hline
$B^- \to D^0 h^-$ & $138.4 \pm 15.5$ & 11.7 & 
$0.079 \pm 0.009 \pm 0.006$ \\
$\bar{B}^0 \to D^+ h^-$ & $33.7 \pm 7.3$ & 6.1 & 
$0.068 \pm 0.015 \pm 0.007$ \\
$B^- \to D^{*0} h^-$ & $32.8 \pm 7.8$ & 5.8 & 
$0.078 \pm 0.019 \pm 0.009$ \\
$\bar{B}^0 \to D^{*+} h^-$ & $36.0 \pm 7.1$ & 7.6 & 
$0.074 \pm 0.015 \pm 0.006$ \\
\hline\hline
\end{tabular} 
\end{center}
\end{table}

\begin{table}[p]
\caption{{}
	Summary of the $B \to \pi\pi$, $K\pi$, $KK$ results.
	The obtained signal yield ($N_s$), statistical significance
	($\Sigma$), reconstruction efficiency ($\epsilon$), charge averaged
	branching fraction ($\cal B$) and its 90\% confidence level upper
	limit (U.L.) are shown.
	In the calculation of ${\cal B}$, the production rates
	of $B^+B^-$ and $B^0\overline{B}^0$ pairs are assumed to be equal.
	In the modes with $K^0$ mesons, $N_s$ and $\epsilon$ are quoted
	for $K^0_S$, while ${\cal B}$ and U.L. are for $K^0$.
	Submode branching fractions for $K^0_S\rightarrow\pi^+\pi^-$
	and $\pi^0\rightarrow\gamma\gamma$ are included in $\epsilon$.
}
\label{tbl:hh1}
\begin{center}
\vspace{0.2cm}
\begin{tabular}{cccccc}
\hline\hline
Mode & $N_s$ & $\Sigma$ &  $\epsilon$ [\%] &
${\cal B}$ [$\times 10^{-5}$] & U.L. [$\times 10^{-5}$]\\ \hline

$B^0\rightarrow\pi^+\pi^-$ & $17.7^{\ +7.1}_{\ -6.4}$ & 3.1 &
28.1 & $0.56^{\ +0.23}_{\ -0.20}\pm 0.04$ & -- \\

$B^+\rightarrow\pi^+\pi^0$ & $10.4^{\ +5.1}_{\ -4.3}$ & 2.7 &
12.0 & $0.78^{\ +0.38\ +0.08}_{\ -0.32\ -0.12}$ & $1.34$ \\

$B^0\rightarrow K^+\pi^-$ & $60.3^{\ +10.6}_{\ -9.9}$ & 7.8 &
28.0 & $1.93^{\ +0.34\ +0.15}_{\ -0.32\ -0.06}$ & -- \\

$B^+\rightarrow K^+\pi^0$ & $34.9^{\ +7.6}_{\ -7.0}$ & 7.2 &
19.2 & $1.63^{\ +0.35\ +0.16}_{\ -0.33\ -0.18}$ & -- \\

$B^+\rightarrow K^0\pi^+$ & $10.3^{\ +4.3}_{\ -3.6}$ & 3.5 &
13.5 & $1.37^{\ +0.57\ +0.19}_{\ -0.48\ -0.18}$ & -- \\

$B^0\rightarrow K^0\pi^0$ & $8.4^{\ +3.8}_{\ -3.1}$ & 3.9 &
9.4 & $1.60^{\ +0.72\ +0.25}_{\ -0.59\ -0.27}$ & -- \\

$B^0\rightarrow K^+K^-$ & $0.2^{\ +3.8}_{\ -0.2}$ & -- & 24.0 & --& 0.27 \\

$B^+\rightarrow K^+\overline{K}{}^0$ & $0.0^{\ +0.9}_{\ -0.0}$ & -- & 12.1 &
-- & 0.50 \\
\hline\hline
\end{tabular}
\end{center}
\end{table}

\begin{figure}[p]
\begin{center}
\mbox{\psfig{figure=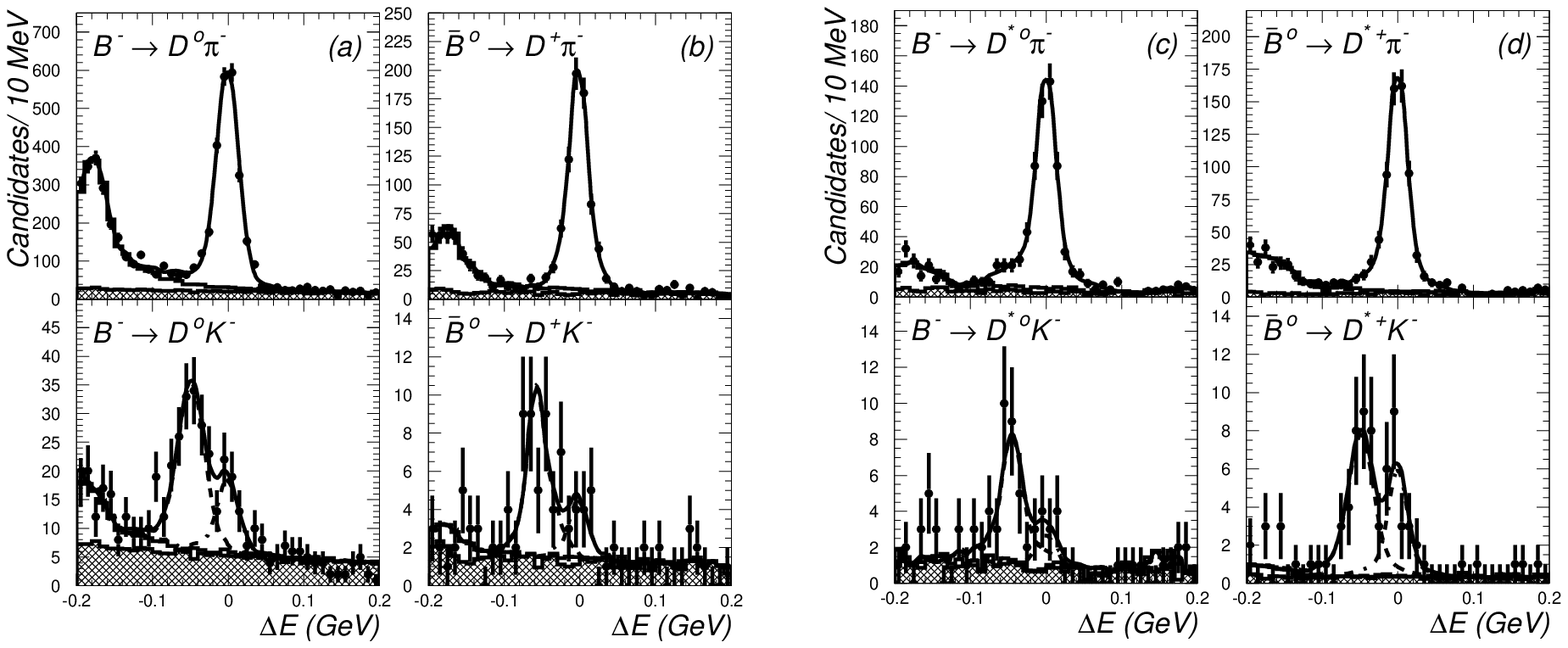,height=6.5cm}}
\end{center}
\caption{
The $\Delta E$ distributions for the
{\bf (a)} $B^-\to D^0 h^-$,
{\bf (b)} $\bar{B}^0\to D^{+} h^-$,
{\bf (c)} $B^-\to D^{*0} h^-$ and
{\bf (d)} $\bar{B}^0\to D^{*+} h^-$
decay channels, in the $m_{bc}$ signal region 
($5.27 \le m_{bc} \le 5.29$\,GeV/$c^2$).
The top figures show $B \to D^{(*)} \pi$ control samples with the PID
cut ${\cal R}_{K} < 0.8$, and the bottom figures show $B \to D^{(*)} K$
enriched samples with the PID cut ${\cal R}_{K} > 0.8$.
The points with error bars present the data, the curves show the
results of fits.
The open histograms are the sums of background functions scaled to
fit the data and the hatched histogram indicates the continuum
component of the background.
}
\label{fig:dk1}
\end{figure}

\begin{figure}[p]
  \begin{center}
  \epsfysize 10.0cm
  \epsfbox{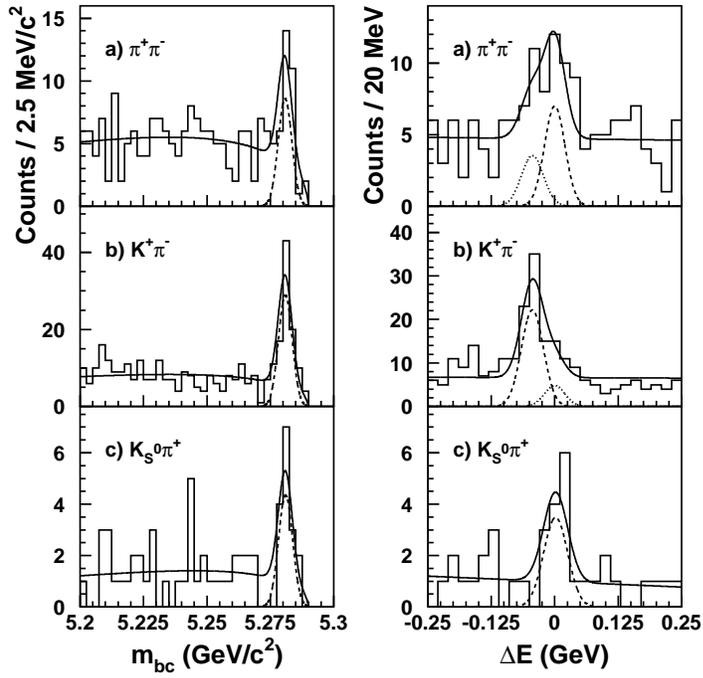}
  \end{center}
  \caption{{}
	The $m_{bc}$ (left) and $\Delta E$ (right) distributions,
	in the signal region of the other variable, for $B\rightarrow$
	a) $\pi^+\pi^-$, b) $K^+\pi^-$ and c) $K^0_S\pi^+$.
	The fit function and its signal component are shown by the solid
	and dashed curves, respectively. 
	In the $\pi^+\pi^-$ and $K^+\pi^-$ fits, the cross talk component
	is shown by the dotted curve.
  }
\label{fig:hh1}
\end{figure}

\begin{figure}[p]
  \begin{center}
  \epsfxsize 10.0cm
  \epsfbox{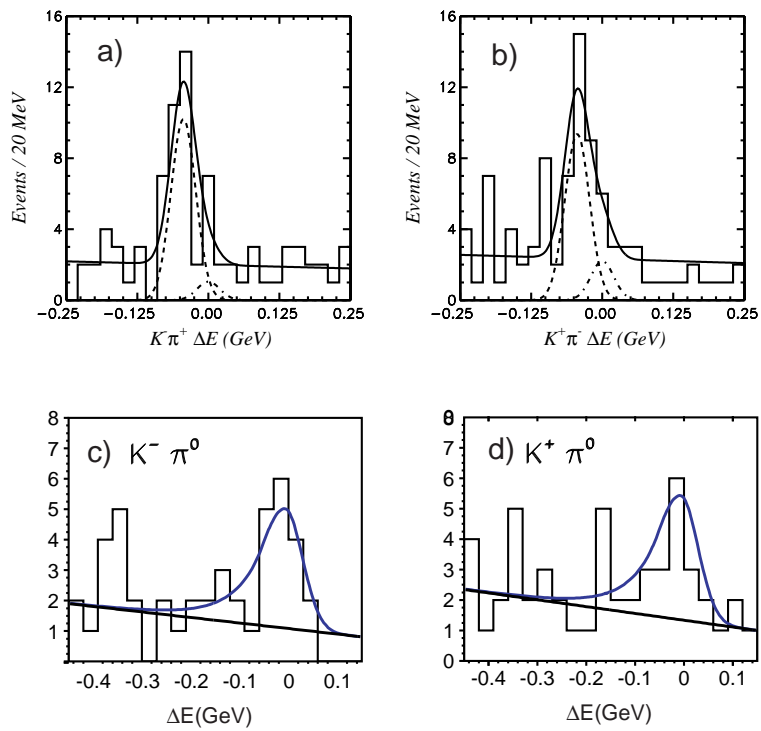}
  \end{center}
\caption{{}
$\Delta E$ distribution for each 
a) $B^0 \to K^- \pi^+$,
b) $\bar{B}^0 \to K^+ \pi^-$,
c) $B^- \to K^- \pi^0$, and
d) $B^+ \to K^+ \pi^0$
mode.
Fits are similar to those shown in Fig.~\ref{fig:hh1} and
Fig.~\ref{fig:hh2}.
}  
\label{fig:hh3}
\end{figure}

\begin{thebibliography}{99}

\bibitem{belle}{
Belle Collaboration, K. Abe {\it et al.},
KEK Progress Report 2000-4 (2000),
to be published in Nucl. Inst. and Meth. A.
}

\bibitem{kekb}{
KEKB B Factory Design Report, KEK Report 95-7 (1995), unpublished.
}

\bibitem{others}{
Belle Collaboration,
A.Bozek, contribution to these proceedings;
A.Garmash, {\it ibid.};
Y.Ushiroda, {\it ibid.};
H.K.Jang, {\it ibid.}.
}

\bibitem{dk:gronau}{
M.Gronau and D.Wyler, \Journal{\PLB}{265}{172}{1991},
I.Dunietz, \Journal{\PLB}{270}{75}{1991},
D.Atwood, I.Dunietz and A.Soni, \Journal{\PRL}{78}{3257}{1997}.
}

\bibitem{dk:cleo}{
CLEO Collaboration, M.Athanas {\it et al.},
\Journal{\PRL}{80}{5493}{1998}.
}

\bibitem{hh:phys}{
See for example:
A.J. Buras and R. Fleischer, Eur. Phys. J. {\bf C16}, 97 (2000);
M. Neubert, Nucl. Phys. Proc. Suppl. 99, 113 (2001);
J. Rosner, in Lecture Notes TASI-2000, World Scientific (2001);
Y.Y. Keum, H.N. Li, A.I. Sanda, \Journal{\PRD}{63}{054008}{2001}.
} 

\bibitem{hh:argus}{
H.Albrecht {\it et al.}, \Journal{\PLB}{241}{278}{1990}.
}

\bibitem{hh:cleo}{
CLEO Collaboration, D.~Cronin-Hennessy {\it et al.},
\Journal{\PRL}{85}{515}{2000}.
} 

\bibitem{hh:babar}{
BABAR Collaboration, T.Champion, Proc. XXXth Int. Conf. on High Energy 
Phys., Osaka, Japan, 2000, edited by C.S.Lim and T.Yamanaka (World
Scientific, Singapore, to be published).
}

\bibitem{etapk:cleo}{
CLEO Collaboration, S.J.Richichi {\it et al.},
\Journal{\PRL}{85}{520}{2000}.
}

\bibitem{kll:vcal}{
CLEO Collaboration, D.M.Asner {\it et al.},
\Journal{\PRD}{53}{1039}{1996}.
}

\bibitem{kll:ali}{
A.Ali, P.Ball, L.T.Handoko and G.Hiller, \Journal{\PRD}{61}{074024}{2000}.
}

\bibitem{kll:cleo}{
CLEO Collaboration, R.Godang {\it et al.}, CLEO CONF 98-22 [in
Proc. XXIX Int. Conf. on High Energy Phys., Vancouver, Canada, 1998,
edited by A.Astbury, D.Axen, and J.Robinson (World Scientific,
Singapore, 1999)].
}

\bibitem{kll:cdf}{
CDF Collaboration, T.Affolder {\it et al.},
\Journal{\PRL}{83}{3378}{1999}.
}

\bibitem{kll:babar}{
BABAR Collaboration, C.Jessop, Proc. XXXth Int. Conf. on High Energy 
Phys., Osaka, Japan, 2000, edited by C.S.Lim and T.Yamanaka (World
Scientific, Singapore, to be published).
}

\end{thebibliography}
\end{document}